\documentclass[twocolumn]{aastex6}

\usepackage{natbib}
\usepackage{amsmath}
\usepackage{multirow}
\usepackage{graphicx}
\usepackage{color}
\usepackage{threeparttable}
\usepackage{enumitem}
\usepackage{mathtools}
\usepackage{lipsum}

\def\maxij1820{MAXI~J1820+070}

\def\swift{{\it Swift}}

\def\nicer{{\it NICER}}

\def\nthComp{{\tt nthComp}}

\def\diskbb{{\tt diskbb}}
\def\powerlaw{{\tt powerlaw}}

\def\reltrans{{\tt reltrans}}
\def\reltransDCp{{\tt reltransDCp}}

\begin{document}
\title{Disk, Corona, Jet Connection in the Intermediate State of \maxij1820\ Revealed by \textit{NICER} Spectral-Timing Analysis}
\author{Jingyi~Wang\altaffilmark{1},
Guglielmo~Mastroserio\altaffilmark{2}, 
Erin~Kara\altaffilmark{1},
Javier~A.~Garc{\'\i}a \altaffilmark{2,3}, 
Adam~Ingram\altaffilmark{4}, 
Riley~Connors\altaffilmark{2}, 
Michiel van der Klis\altaffilmark{5}, 
Thomas~Dauser\altaffilmark{3}, 
James~F.~Steiner\altaffilmark{6},
Douglas~J.~K.~Buisson\altaffilmark{7,8},
Jeroen~Homan \altaffilmark{9,10},
Matteo~Lucchini\altaffilmark{1,5}, 
Andrew~C.~Fabian\altaffilmark{8},
Joe~Bright\altaffilmark{4}, 
Rob~Fender\altaffilmark{4}, 
Edward~M.~Cackett\altaffilmark{11}, 
Ron~A.~Remillard\altaffilmark{1}.
}
\affil{
\altaffilmark{1}MIT Kavli Institute for Astrophysics and Space
Research, MIT, 70 Vassar Street, Cambridge, MA 02139, USA\\
\altaffilmark{2}Cahill Center for Astronomy and Astrophysics, California Institute of Technology, Pasadena, CA 91125, USA\\ 
\altaffilmark{3}Remeis Observatory \& ECAP, Universit\"{a}t
Erlangen-N\"{u}rnberg, 96049 Bamberg, Germany\\ 
\altaffilmark{4}Department of Physics, Astrophysics, University of Oxford, Denys Wilkinson Building, Keble Road, Oxford OX1 3RH, UK\\
\altaffilmark{5}Astronomical Institute, Anton Pannekoek, University of Amsterdam, Science Park 904, NL-1098 XH Amsterdam, Netherlands\\
\altaffilmark{6}Harvard-Smithsonian Center for Astrophysics, 60 Garden St., Cambridge, MA 02138, USA \\
\altaffilmark{7}Department of Physics and Astronomy, University of Southampton, Highfield, Southampton, SO17 1BJ\\
\altaffilmark{8}Institute of Astronomy, University of Cambridge, Madingley Road, Cambridge, CB3 0HA\\
\altaffilmark{9}Eureka Scientific, Inc., 2452 Delmer Street, Oakland, CA 94602, USA\\
\altaffilmark{10}SRON, Netherlands Institute for Space Research, Sorbonnelaan 2, 3584 CA Utrecht, The Netherlands\\
\altaffilmark{11}Department of Physics \& Astronomy, Wayne State University, 666 W. Hancock St, Detroit, MI 48201, USA\\
}

\begin{abstract}
We analyze 5 epochs of \nicer\ data of the black hole X-ray binary \maxij1820\ during the bright hard-to-soft state transition in its 2018 outburst with both reflection spectroscopy and Fourier-resolved timing analysis. We confirm the previous discovery of reverberation lags in the hard state, and find that the frequency range where the (soft) reverberation lag dominates decreases with the reverberation lag amplitude increasing during the transition, suggesting an increasing X-ray emitting region, possibly due to an expanding corona. By jointly fitting the lag-energy spectra in a number of broad frequency ranges with the reverberation model \reltrans, we find the increase in reverberation lag is best described by an increase in the X-ray coronal height. This result, along with the finding that the corona contracts in the hard state, suggests a close relationship between spatial extent of the X-ray corona and the radio jet. We find the corona expansion (as probed by reverberation) precedes a radio flare by $\sim 5$~days, which may suggest that the hard-to-soft transition is marked by the corona expanding vertically and launching a jet knot that propagates along the jet stream at relativistic velocities.
\end{abstract}

\section{Introduction} \label{intro}
There are two populations of astrophysical black holes with strong observational evidence: stellar-mass black holes and supermassive black holes. Despite the orders-of-magnitude black hole mass difference, the two populations share many spectral and timing features, particularly in the X-ray band \citep{merloni2003fundamental, falcke2004scheme, uttley2014x}. On the other hand, they also have distinct astrophysical significance. In the case of stellar-mass black holes, the distribution of spins and masses measured in black hole binary (BHB) systems observed in X-rays, and in those of mergers involving black holes detected via gravitational waves by LIGO/Virgo, will shed light on their formation channels \citep{reynolds2020observational}. The supermassive black holes in active galactic nuclei (AGN) are believed to profoundly influence galaxy evolution via the radiation and outflows, which is referred to as AGN feedback \citep{fabian2012observational}. 

\begin{figure}[htb!]
\centering
\includegraphics[width=\linewidth]{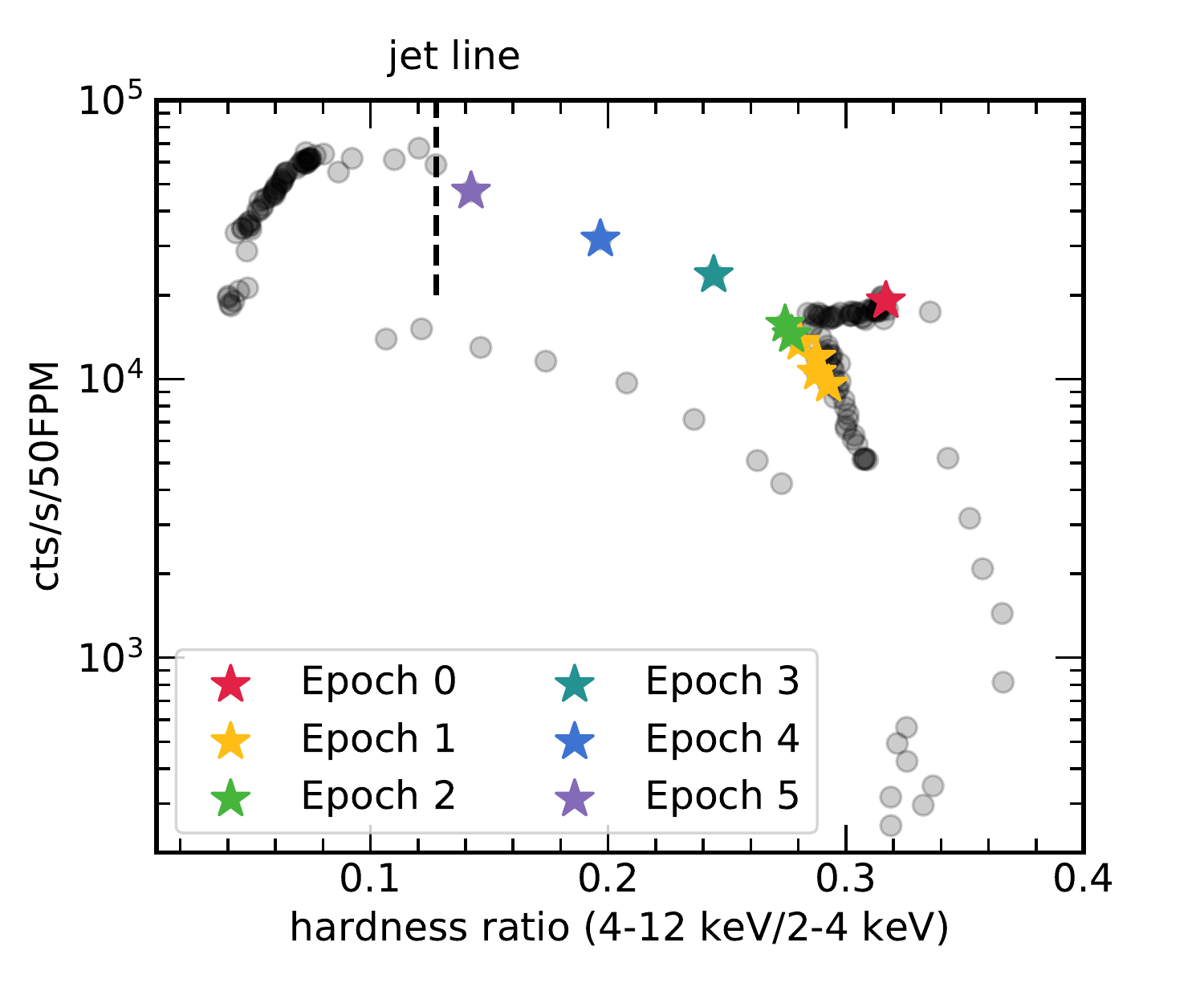}
\caption{
The \nicer\ hardness-intensity diagram (HID) where the hardness ratio is defined as the count ratio of the hard band (4--12~keV) and the soft band (2--4~keV), and the count rate is in 0.4--12~keV and normalized for 50 FPMs. The grey points represent the \nicer\ observations from 2018 March 6 to 2018 Oct 15, corresponding to obs.ID 1200120101 to 1200120278. Some points appear black because of several observations with very similar hardness and count rate overlapping. The ``jet line" corresponds to the QPO type transition and the observed strong radio flare \citep{homan2020rapid}. 
}
\label{fig:hid}
\end{figure}

BHBs exhibit a variety of accretion states with rich multi-wavelength characteristics in a single system, on a human timescale \citep{remillard2006x}. The outburst of BHBs can be described by a hysteresis pattern in the hardness-intensity diagram (HID; e.g., \citealp{belloni2016transient}, see also Fig.~\ref{fig:hid}). In a complete outburst, the BHB rises from the quiescent state when the accretion rate is very low, usually $\lesssim10^{-5}$ of the Eddington limit, to the hard state when the X-ray emission is dominated by the non-thermal coronal emission. The luminosity keeps rising until it makes a transition into the soft state dominated by the thermal disk emission, and the transitional state is called the intermediate state. Then, with a steadily decreasing mass accretion rate, the BHB eventually transitions back to hard state and fades into quiescence. In the radio band, a compact and mildly relativistic jet is commonly observed in the hard state (e.g. \citealt{dhawan2000scale}). During the hard-to-soft transition, the core jet is slowly quenched, and at the so-called ``jet line", which loosely corresponds to the boundary of the hard and soft intermediate states, a highly relativistic, discrete and ballistic jet is launched with unclear mechanism \citep{fender2004towards}. 

The X-ray emission from a BHB system consists of the thermal emission from the accretion disk, the ubiquitous non-thermal continuum emission caused by Compton upscattering of disk photons intercepted by hot (of the order of $\sim100$~keV) electrons in the ``corona", and a reflection component as the disk material reprocesses the illuminating coronal photons. The local reflection spectrum is composed of prominent features such as the Fe K emission complex ($6-7$~keV), the Fe K-edge ($\sim 7-10$\,keV), and the Compton hump ($\sim 20-30$\,keV), and it is smeared by relativistic effects when produced close to the central object \citep{fabian1989x}. Within this paradigm, a type of time lag feature called a reverberation lag is naturally anticipated because of the light travel time delay between the direct coronal and the reflected emission. The reverberation lag has been detected in both AGN and BHBs, in the forms of the soft-excess emission and/or the Fe K emission line lagging behind the continuum-dominated band (e.g., \citealp{fabian2009broad, kara2016global, de2017evolution}). 

There are many remarkable questions waiting to be answered which are significant to understanding accretion and ejection physics in the strong gravity regime, such as the nature of the corona, and the state transition mechanism. Despite the ubiquity of the non-thermal X-ray emission, the fundamental properties of the corona are not well understood \citep{done2007modelling}. The difficulty lies in the fact that the innermost region of BHBs are too compact to be resolved via imaging, and it is challenging for X-ray spectroscopy alone to distinguish the underlying continuum models (e.g., single or double Comptonization regions; \citealp{dzielak2019comparison}) from which the geometry and formation of the corona could then be inferred. There are three different types of coronal models: radiatively inefficient accretion flow (RIAF) within the truncated accretion disk (advection-dominated accretion flow or other types of coronal flow; \citealp{narayan2008advection, ferreira2006unified}), static corona with different configurations (e.g., slab; \citealp{haardt1994model}), and the corona being the base of a jet \citep{markoff2001jet,markoff2005going}. One commonly adopted geometry is the lamppost geometry, which could very likely be due to the base of a jet, is also motivated by relativistic reflection itself in terms of the steep emissivity profile \citep{dauser2013irradiation,wilkins2012understanding}.

The hypothesis that the corona is the base of a jet is motivated by observational evidence. The connection between the X-ray corona and the radio-emitting jet was first observed in the BHB GX~399$-$4 \citep{hannikainen1998most,corbel2000coupling}, and now is seen universally in the hard state of low mass X-ray binaries (LMXBs; \citealp{gallo2003universal, fender2004towards, gallo2012assessing}). Strong correlations between optical/NIR and X-ray observations also indicate a connection between the corona and the jet \citep{homan2005multiwavelength,russell2006global,gandhi2007}. This relation can even be extended to AGN, in which it is called the ``fundamental plane of black hole activity", and shows that there is a connection between the inner accretion flow (as probed by X-rays) and large scale jets seen in the radio \citep{merloni2003fundamental,falcke2004scheme,plotkin2012using}. Such observations led to models that suggest that the corona is in fact the base of a relativistic jet, where the hard X-rays are produced by inverse Compton scattering of thermal disc photons and synchrotron photons in the jet base \citep{fender1999quenching,markoff2005going}. Such a model, fitted to the broadband SED from radio to hard X-ray, suggests a corona/jet-base outflow \citep{markoff2004constraining}. Similar models have been extended to higher luminosities and larger masses to explain the reflection and reverberation-constrained coronal geometry in AGN \citep{king2017agn}.

The state transition mechanism is also a remaining puzzle with challenges arising from both theoretical and observational perspectives. In the theory side, it is computationally expensive to simulate long term behaviors of accretion flows to track through accretion states; while in the observational arena the hard-to-soft transition happens on a fairly fast timescale of usually a week, so the datasets are valuable. It has been proposed that the inner hot flow collapses and transitions into a standard optically thick accretion disk during the hard-to-soft transition. However, this point of view is challenged by measurements of the inner disk radius, $R_{\rm in}$, being within $\sim10 R_{\rm g}$ ($R_{\rm g}=GM_{\rm BH}/c^2$) in the bright hard state (above 1\% of the Eddington limit; \citealp{javier_gx339, wang2018evolution}), and is persistent at $<9R_{\rm g}$ during the transition \citep{sridhar2020evolution}.

\maxij1820 is one of the brightest BHBs detected so far, and has exhibited a great amount of activity in radio \citep{homan2020rapid, bright2020extremely}, near infrared \citep{sanchez2020near}, optical \citep{veledina2019evolving, shidatsu2019x, munoz2019hard, paice2019black}, UV \citep{kajava2019x}, and X-ray \citep{ kara2019corona,buisson2019maxi, espinasse2020relativistic, wang2020evolution, buisson2021maxi,you2021,zdziarski2021accretion}. In the hard state, reverberation analysis using the \textit{Neutron Star Interior Composition Interior Explorer (NICER}, \citealp{gendreau2016neutron}) showed that while the broad iron line profile remained constant (and only the narrow component decreased), the frequency range where the (soft) reverberation lag dominated increased \citep{kara2019corona}. This result was interpreted as a contraction in the X-ray corona. This hypothesis has been later confirmed by reflection spectroscopy in the context of a dual-lamppost, where the height of the upper lamppost decreased while the lower one stayed constant \citep{buisson2019maxi}, and by the increase of the peak frequency of the quasi periodic oscillations \citep[QPO;][]{ingram2019review}. During the hard-to-soft transition, an extremely powerful superluminal ejection was observed in the radio band \citep{bright2020extremely} very close in time with a small flare in the 7--12 keV band and a transition from type-C to type-B QPO \citep{homan2020rapid}. 

While reflection spectroscopy and reverberation mapping fundamentally probe the same region around the black hole, until very recently, the spectral and timing modeling were done independently. Spectral-timing modeling together with the vast improvement in high-throughput data mean that we have the potential for studying the inner accretion flow with unprecedented detail. 
One model with this capability is \reltrans\ \citep{ingram2019public}. In \reltrans, a lamppost corona and razor thin disk are assumed, where the lamppost is the simplest prescription of the corona as the base of the jet. There are two main kinds of time lags: the low-frequency hard lags, which are fairly ubiquitous in both AGN and BHBs (e.g., \citealp{van1987complex,nowak1999rossi,papadakis2001frequency}), and are described as fluctuations in the photon index $\Gamma$ of the illuminating coronal emission; and the high-frequency soft/reverberation lags, which are caused by the light-crossing time delay between the continuum and reflected emission. Non-linear changes in the reflection spectrum due to the variable continuum spectrum are approximated via first-order Taylor expansion \citep{mastroserio2018multi}. The model can be used to fit directly and simultaneously to both the real and imaginary parts of the cross-spectrum as a function of energy in different frequency bands, accounting correctly for the instrumental response, and self-consistently produces the time-averaged energy spectrum. It has led to successes in fitting to observational data of both AGN \citep[Mrk~335;][]{mastroserio2020multi} and BHB systems \citep[Cyg~X-1;][]{mastroserio2019x}. 

The \nicer\ dataset of \maxij1820\ is exceptional because it covers the bright state transition which happens on a short timescale. It was so bright that we can extend previous reverberation lag analysis into the state transition, which is usually difficult because of the decreased variability and limited signal-to-noise ratios of lag spectra. 

In this paper, we first qualitatively explore the evolution of iron line profile and the reverberation lag frequency and amplitude, and then fit the lag-energy spectra and time-averaged energy spectrum with the physically motivated reverberation model \reltrans. This paper is organized as follows. Section~2 describes the observations and data reduction, Section~3 provides the details of results, including a phenomenological first look (Section~3.1), the lag-energy spectral fitting (Section~3.2), and the time-averaged spectral fitting (Section~3.3). We present the discussion in Section~4, and summarize the results in Section~5.

\begin{table*}[htb!]
\begin{center}
\caption{\nicer\ observations in the 2018 outburst
cycles, exposure times and dates. \label{tab:obs}}
\footnotesize
\begin{tabular}{cccccccc}\hline \hline
State&Epoch&obs.ID$^\dagger$&Date&MJD&exp.(ks)&counts/s/50FPMs&hardness ratio\\
\hline
hard&0&106&03/21&58198&5.4&17750&0.32\\
\hline
hard-to-soft transition&1&189-192&06/28-07/01&58297-58300&19.4&9055&0.29\\
&2&193-194&07/02-07/03&58301&6.0&14190&0.28\\
&3&195&07/04&58303&0.7&21730&0.24\\
&4&196&07/05&58304&5.9&25650&0.20\\
&5&197 (before flare)&07/06&58305&13.0&44740&0.14\\
\hline
\end{tabular}
\\
\raggedright{\textbf{Notes.} \\The count rates are for $0.5-10$~keV, and the hardness ratio is defined as the count ratio of the hard band (4-12~keV) and the soft band (2-4~keV). $^\dagger$ObsIDs for \nicer\ are 1200120xxx. }
\end{center}
\end{table*}

\section{Observations and data reduction} \label{obs}
\maxij1820 is also known as ASASSN-18ey since it was discovered in optical by the All-sky Automated survey for supernovae (ASAS-SN, \citealp{shappee2014man}) on 2018 March 6 \citep{tucker2018asassn}, and was only discovered in X-ray by the Monitor of All-sky X-ray Image (MAXI, \citealp{matsuoka2009maxi}) 5~days later \citep{kawamuro2018maxi}.  
\nicer\ monitored it on a daily cadence, from 2018 March 6 to 2018 November 21, making it an extraordinary dataset covering the bright hard-to-soft transition which usually happens fast ($\sim$ a week) and thus is difficult to catch. The \nicer\ HID is shown in Fig.\ref{fig:hid}, where Epochs 1-5 were chosen to cover the hard-to-soft transition are shown, along with Epoch 0 serving as a reference point in the bright hard state (the same as the first epoch investigated in \citealp{kara2019corona}). In Epochs 1 and 2, we combine several observations that are close in time and hardness ratio, to boost the signal-to-noise ratios (see Tab.~\ref{tab:obs}, also for details in other epochs). We note that Epoch~5 (obs.ID 1200120197) corresponds to the X-ray flare seen in \citet{homan2020rapid}, when type-C QPOs transitions into type-B, and a strong radio flare was observed $\sim2$~hours after that (see the ``jet line" in Fig.\ref{fig:hid}). To avoid the type-B QPO, we select the photons with arrival time before MJD 58,305.67740 (TIME$< 142359302$) in that observation. 

The \nicer\ data are processed with the data-analysis software NICERDAS version 2019-05-21\_V006, and CALDB version xti20200722 with the energy scale (gain) version ``optmv10". We use the standard filtering criteria: the pointing offset is less than $54\arcsec$, the pointing direction is more than $40^\circ$ away from the bright Earth limb, more than $30^\circ$ away from the dark Earth limb, outside the South Atlantic Anomaly (SAA). We also filter out the events from the two noisy detectors, FPMs 34 and 14. In addition, we select events that are not flagged as ``overshoot" or ``undershoot" resets (EVENT\_FLAGS=bxxxx00), or forced triggers (EVENT\_FLAGS=bx1x000). A ``trumpet" filter is also applied to filter out known background events \citep{bogdanov2019constraining}. The cleaned events are barycenter corrected using the \texttt{FTOOL} \texttt{barycorr}. The spectra are then binned with a minimum count of 25 per channel, and the energy resolution was oversampled by a factor of 3. We do not include background in the analysis because of the very high count rate of the source, and the fitted energy range is 0.5-10~keV. We use the RMF version ``rmf6s" and ARF version ``consim135p", which are both a part of the CALDB xti20200722, and we combine 50 modules excluding FPMs 34 and 14. We note that some detectors were turned  off starting from ObsID 1200120196 due to the high count rate, so the response matrices for those observations are combined differently and accordingly. For the purpose of timing analysis, the light curve segment length is 10~s with 0.001~s bins, which covers frequencies from 0.1 to 500~Hz. 

All the uncertainties quoted in this paper are for a 90\%
confidence range, unless otherwise stated. All spectral fitting is done with XSPEC 12.11.1 \citep{arn96}. In all of the fits, we use the \textit{wilm} set of abundances \citep{wilms2000}, and \textit{vern} photoelectric cross sections \citep{verner1996atomic}. We used the $\chi^2$ fit statistics.

\section{Results} \label{results}

\begin{figure*}[htb!]
\centering
\includegraphics[width=0.8\linewidth]{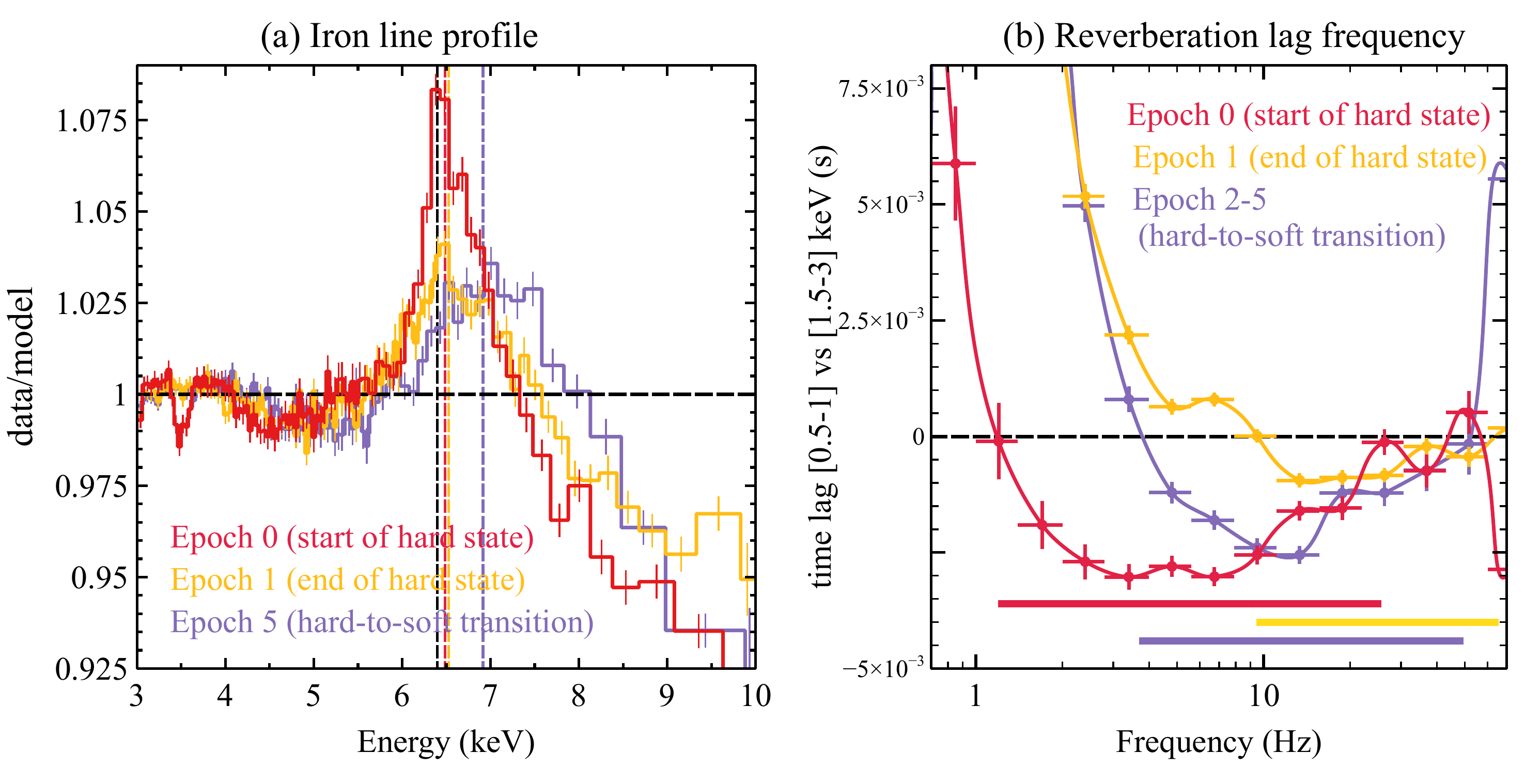}
\caption{The evolution of iron line profile and reverberation lag frequency. 
(\textit{a}) The data-to-model ratios when \nicer\ spectra are fitted in 3--10~keV range with a model of (\diskbb+\powerlaw). The vertical lines are at 6.4~keV (black), and the fitted \texttt{gaussian} line energies of 6.48~keV (Epoch~0), 6.52~keV (Epoch~1), and 6.92~keV (Epoch~5). The color of each epoch matches that in Fig.~\ref{fig:hid}. (\textit{b}) The lag-frequency spectra. The logarithmic frequency rebinning factor is 0.4, and we use the convention that a positive lag indicates a hard lag, i.e., the hard photons lag behind the soft ones. The horizontal lines represent the frequency ranges of soft lags. The soft lag frequency increases in the hard state and then decreases during the transition. We show the combined lag-frequency spectrum in the transition (Epochs~2--5) for clarity; the individual epochs exhibit the same decrease of soft lag frequency compared to the end of hard state (Fig.~\ref{fig:individual_tlag}).
}
\label{fig:iron_line}
\end{figure*}

\subsection{A phenomenological first look} \label{phenomena}
We first fit the time-averaged energy spectra in 3--10~keV with a model of (\diskbb+\powerlaw), and the data-to-model ratios are shown in Fig.~\ref{fig:iron_line}~(a), in order to qualitatively asses the excess flux in the iron K$\alpha$ region. A line-like feature exists in all epochs, which clearly evolves through the epochs. In the hard state (Epochs~0--1), it remains stable, while the strength of the narrow component decreases, as shown in \citet{kara2019corona} and \citet{buisson2019maxi}. During the hard-to-soft transition, the line feature remains very similar until Epoch~5, when the line gets blue shifted. When an extra \texttt{gaussian} component is added to approximate the iron K emission, the line energy increases from $6.52\pm0.05$~keV (Epoch~1) to $6.92\pm0.05$~keV (Epoch~5), with tied line width. We explore models for the evolution of the reflection features in Section~\ref{spec_fit}. 

From the timing perspective, we calculate the cross-spectrum in each epoch following standard Fourier timing techniques \citep{uttley2014x}, between the energy bands of $0.5-1$~keV and $1.5-3$~keV. We show in Fig.~\ref{fig:iron_line}~(b) the lags as a function of frequency for Epoch~0 and 1, representing the beginning and end of the hard state, and also for the combined Epochs~2--5 due to the significantly reduced variability during the transition. All epochs show hard lags at low frequencies and soft lags at high frequencies, but the frequency range where the soft lag dominates is evolving. In the hard state, we confirm the results from \citet{kara2019corona} that the soft lag frequency increases, along with a decreasing lag amplitude; while in the transition, we find the soft lag shows an opposite trend, when its frequency decreases and the amplitude increases. The behavior in the transition suggests a growing emitting region, possibly due to an expansion of the corona. We show the combined lag-frequency spectrum (Epochs~2--5) here for clarity; the individual epochs (see Fig.~\ref{fig:individual_tlag}) exhibit the trend in more detail which is also confirmed in the lag-energy spectral fits (see Section~\ref{lag_fit}).

\subsection{Time-averaged flux spectral fitting} \label{spec_fit}
As shown in Section~\ref{phenomena} and Fig.~\ref{fig:iron_line}, a reflection feature is present in the time-averaged energy spectra of all epochs. We first perform a simultaneous fit to Epochs~1-5 with the model \texttt{(tbabs*diskbb+reltransDCp)} where \texttt{reltransDCp}\footnote{We note that the {\tt reltransDCp} model used here implements the {\tt xillverDCp} reflection models, which include a Comptonization continuum and variable density. These models ({\tt xillverDCp, reltransDCp}) are not public yet, but they will be released in upcoming updates. The model returns the time-averaged energy spectrum (by setting parameter $f_{\rm min}=f_{\rm max}=0$). We use a radial ionization profile self-consistently calculated from the lamppost illumination (20 radial zones in the disk are used with \texttt{ION\_ZONES}=20), and account for a radial density profile following the zone A of the \citet{shakura1973black} disk model (\texttt{A\_DENSITY}=1). The galactic absorption is self-consistently accounted for in \reltransDCp, and $N_{\rm H}$ are tied between the one in \reltransDCp\ and the one acting on \diskbb.} is a flavor of \reltrans, which adopts a more physically motivated thermal Comptonization model \nthComp\ \citep{zdziarski1996broad} as the illuminating coronal emission, and can also probe high electron density ($N_{\rm e}$) up to $10^{20}$~cm$^{-3}$ rather than a fixed value of $10^{15}$~cm$^{-3}$. The high density model has been shown to contribute to soft excess emission below 2~keV \citep{garcia2016effects} and may have the effect of decreasing the fitted iron abundance, which is usually found to be super-solar \citep{garcia2018_afe,jiang2019high}. 

\begin{figure*}[htb!]
\centering
\includegraphics[width=1.\linewidth]{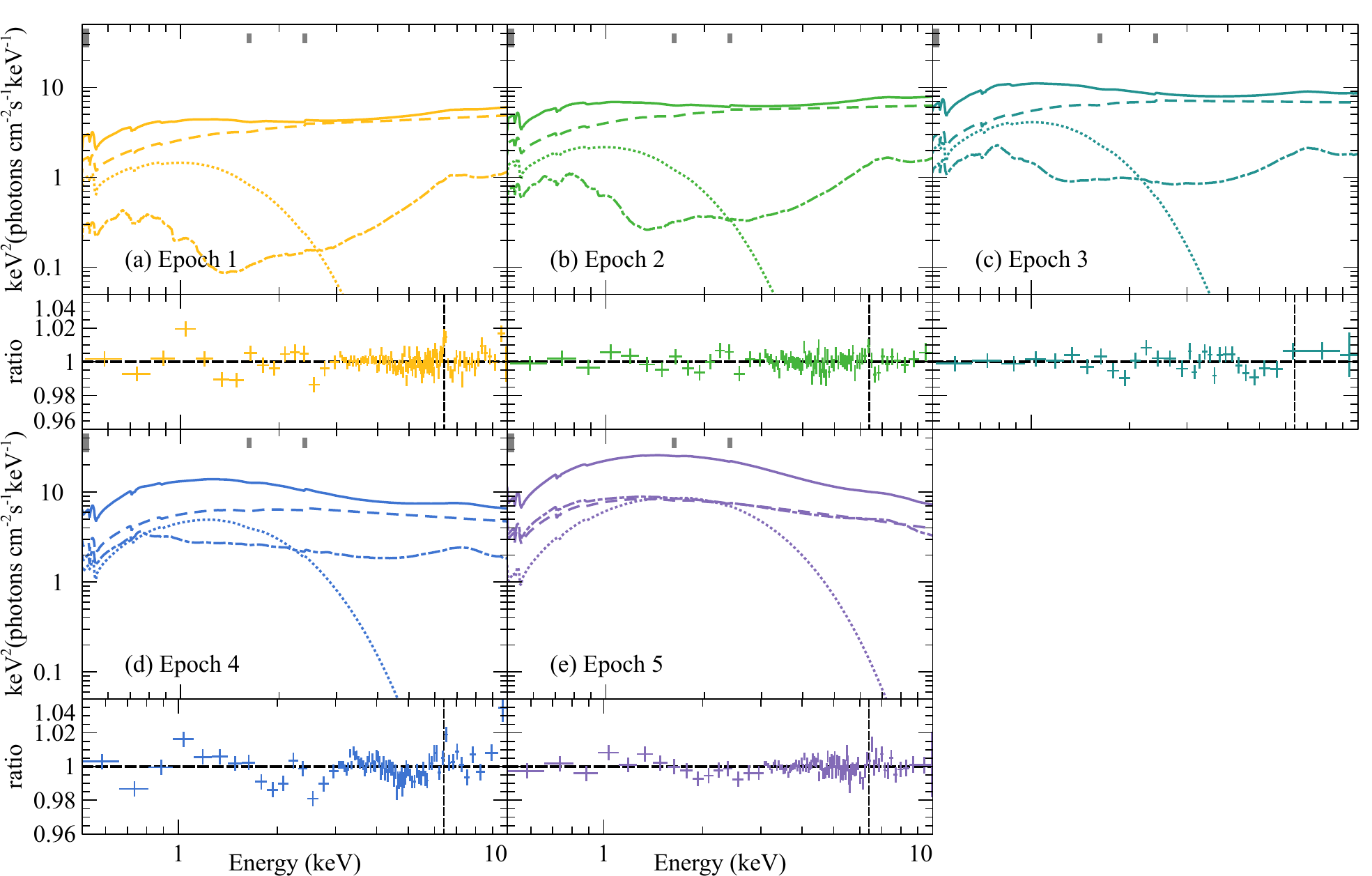}
\caption{
The model (\textit{upper}) and data-to-model ratios (\textit{lower}) when \nicer\ spectra are fitted in 0.5-10~keV range with a model of \texttt{(tbabs*diskbb+reltransDCp)*edge*edge*gabs}. The total model, the \texttt{diskbb}, the illuminating continuum, and the reflection component are represented by solid, dotted, dashed, and dash-dotted lines respectively. The continuum and reflection overlap in Epoch~5 due to a boost parameter of $3.0\pm1.5$, where the best-fit value is larger than previous epochs, but becomes consistent within 90\% confidence. The vertical line is at 6.4~keV, and the color of each epoch matches that in Fig.~\ref{fig:hid}. In Epochs~1 and 2, the residual is very narrow and peaks precisely at 6.4~keV, and can be modeled with an additional \texttt{xillverDCp} component; while in Epochs~4 and 5 (when the source approaches the jet line), the residual is excess above 6.4~keV (i.e., more blue shifted). See the Section~\ref{spec_fit} for more details. The vertical grey bars indicate the characteristic energies of phenomenological models.
}
\label{fig:eem_ratio}
\end{figure*}

The global parameters including the column density, inclination, and iron abundance are all tied among epochs, and a maximal spin is assumed ($a_*=0.998$) to fit for $R_{\rm in}$. We also add a 1\% systematics to energies below 3~keV to reduce the effects from calibration uncertainties at low energies, and at the same time raise the statistical importance of the iron line band. The resulting $\chi^2$/d.o.f. is $3169/1112=2.85$, with strong residual features at low energies $<3$~keV (see the lower panel of Fig.~\ref{fig:cal_model} for the contributions to $\chi^2$ in Epoch~3 as a representative case). The most noticeable features include edge-like shapes near $\sim0.5$~keV and $\sim2.4$~keV, and Gaussian-like absorption around $1.6$~keV. These remaining features are likely from \nicer's calibration systematics because they could be attributed to oxygen ($\sim0.5$~keV), aluminum K edge ($\sim1.56$~keV), and gold M edge ($2.1-4.5$~keV complex), and thus we model them empirically. 

Similar to \citet{wang2020relativistic}, we adopt phenomenological models (\texttt{edge} and \texttt{gabs}) to model those features empirically. This leads to the final model in this work: 
\begin{equation}
\begin{multlined}
	\texttt{(tbabs*diskbb+reltransDCp)}\texttt{*edge*edge*gabs}. \nonumber
\end{multlined}
\end{equation}

In the phenomenological models, only their energies are tied through epochs, allowing
for possible attitude-dependent variations which are not accounted for in the current response matrix. The fit with the phenomenological calibration model has $\chi^2$/d.o.f.=1313.8/1089=1.21, decreasing significantly by 1855 with 23 fewer d.o.f. compared to the model without calibration features. The contributions from model components, and the data-to-model ratios are shown in Fig.~\ref{fig:eem_ratio}. The best-fit parameters are presented in Table~\ref{tab:spec_pars}. We show the effective area of \nicer\ in Fig.~\ref{fig:cal_model} (upper), with the vertical lines indicating the fitted characteristic energies of phenomenological models, in this work and in a previous \nicer\ spectral analysis of another BHB GX~339--4 \citep{wang2020relativistic}. We notice the characteristic energies are not identical, and that could be due to the updated gain solution and response matrices; and also in previous work, a Crab correction was applied. We suggest readers be aware of such residuals and follow the most recent \nicer\ analysis guidelines. 

\begin{table*}[htb!]
\caption{Best fit parameters when fitting the time-averaged spectra with the model \texttt{(tbabs*diskbb+reltransDCp)*edge*edge*gabs}.  \label{tab:spec_pars}}
\centering
\begin{tabular}{ccccccc}\hline \hline
Parameter & Epoch 1&Epoch 2 &Epoch 3&Epoch 4&Epoch 5\\
\hline
$N_{\rm H}$&\multicolumn{5}{c}{$(1.73^{+0.10}_{-0.07})\times 10^{21}$cm$^{-2}$}\\
$a_*$&\multicolumn{5}{c}{$0.998$}\\
$i$ (degrees)&\multicolumn{5}{c}{$81.0^{(p)}_{-0.1}$ }\\
$A_{\rm Fe}$&\multicolumn{5}{c}{$1.2^{+0.1}_{-0.2}$}\\
$E_{\rm edge,1}$ (keV)&\multicolumn{5}{c}{$0.5000^{+0.0002}_{(p)}$}\\
$E_{\rm edge,2}$ (keV)&\multicolumn{5}{c}{$2.40^{+0.02}_{-0.01}$}\\
$E_{\rm gabs}$ (keV)&\multicolumn{5}{c}{$1.62^{+0.01}_{-0.02}$}\\
$T_{\rm in}$ (keV)& $0.318^{+0.005}_{-0.004}$ & $0.314^{+0.006}_{-0.005}$ & $0.311^{+0.006}_{-0.005}$ & $0.406\pm0.004$ & $0.601^{+0.004}_{-0.006}$\\
${\rm N}_{\rm diskbb}$ ($10^4$)& $3.0^{+0.2}_{-0.3}$ & $4.7\pm0.4$ & $9.6^{+0.7}_{-0.8}$ & $3.5\pm0.2$ & $1.12^{+0.06}_{-0.04}$\\
$R_{\rm in}$ ($R_{\rm ISCO}$) & $<1.2$ & $1.3\pm0.2$ & $<1.1$ & $1.6^{+0.1}_{-0.2}$ & $7.5^{+0.5}_{-0.9}$ \\
$\log N_{\rm e}$ (cm$^{-3}$) & $15.9\pm0.1$ & $15.8^{+0.2}_{-0.3}$ & $16.8^{+0.3}_{-0.5}$ & $16.2\pm0.2$ & $<15.5$ \\
$\Gamma$& $1.84\pm0.01$ & $1.91\pm0.02$ & $2.03^{+0.03}_{-0.02}$ & $2.21\pm0.01$ & $2.42^{+0.03}_{-0.01}$ \\
$kT_{\rm e}$ (keV)&$<74$  & $<217$ & $>31$ & $>136$ & $<34$\\
$h$ ($R_{\rm g}$)&  $25\pm2$ & $25\pm2$ & $22^{+4}_{-2}$ & $20\pm2$ & $9^{+2}_{-1}$\\
$\log \xi$ (erg$\cdot$cm$\cdot$s$^{-1}$) & $2.48^{+0.10}_{-0.17}$ & $2.47^{+0.23}_{-0.25}$ & $3.07^{+0.22}_{-0.80}$ & $3.48^{+0.09}_{-0.11}$ & $4.54^{+0.07}_{-0.14}$ \\
boost & $1.8\pm0.2$ & $2.1^{+0.2}_{-0.3}$ & $1.7^{+0.4}_{-0.2}$ & $1.7\pm0.1$ & $3.0\pm1.5$ \\
N$_{\rm r}$ ($10^{-2}$)& $7.0^{+0.5}_{-0.2}$ & $8.8^{+1.9}_{-0.6}$ & $12.2^{+0.4}_{-0.6}$ & $11.0^{+0.4}_{-0.8}$ & $19.7^{+1.2}_{-1.8}$ \\
$\tau_{\rm max,1}$ & $-0.44^{+0.09}_{-0.06}$ & $-0.37^{+0.06}_{-0.09}$ & $-0.0\pm0.08$ & $0.24^{+0.06}_{-0.09}$ &  $0.56^{+0.07}_{-0.08}$ \\
$\tau_{\rm max,2}$ & $-0.055^{+0.006}_{-0.005}$ & $-0.049\pm0.007$ & $-0.04\pm0.01$ & $-0.059^{+0.009}_{-0.007}$ &  $-0.019\pm0.006$ \\
$\sigma$ (keV)& $>0.09$ & $>0.09$ & $>0.09$ & $>0.09$ & $0.08^{(p)}_{-0.03}$ \\
Strength & $0.012\pm0.002$ & $0.011\pm0.002$ & $0.011\pm0.002$ & $0.012\pm0.002$ & $0.004\pm0.002$ \\
$\chi^2$/d.o.f.& \multicolumn{5}{c}{$1349.2/1089=1.24$}\\
\hline
\end{tabular}
\\
\raggedright{\textbf{Notes.} \\
Errors are at 90\% confidence level and statistical only. The inclination is limited to be $<81^\circ$ given by the binary inclination \citep{torres2020binary}. For the phenomenological model \texttt{gabs}, the width is set to have an upper limit of 0.1~keV, and it is pegged at that upper limit in all epochs. Parameters without uncertainties were fixed at the quoted value. The index $(p)$ means that the uncertainty is pegged at the bound allowed for the parameter.}
\end{table*}

In Fig.~\ref{fig:eem_ratio}, we see this basic model is insufficient, and the shapes of residuals at the iron line are different among the epochs. At the end of hard state (Epoch~1) and the beginning of transition (Epoch~2), the residual is very narrow and peaks precisely at 6.4~keV, which has been seen previously with reflection spectroscopy in the hard state, and is usually interpreted as a distant reflector (e.g., \citealt{javier_gx339}). Moreover, the narrow line has been seen in \maxij1820\ \citep{buisson2019maxi}. If an additional \texttt{xillverDCp} component is added in Epochs~1 and 2, with the \texttt{xillverDCp} parameters tied to the corresponding ones in \texttt{reltransDCp} except for the ionization parameter and the normalization, $\chi^2$ decreases by 38 (Epoch~1) and 10 (Epoch~2) each with 2 fewer d.o.f., and the key parameter values shown in Table~\ref{tab:spec_pars} stay similar.

However, when the source approaches the jet line, the residuals in Epochs~4 and 5 are more blue shifted (i.e., excess above $6.4$~keV, instead of a narrow peak at $6.4$~keV), as already indicated by Fig.~\ref{fig:iron_line}. This behavior makes the modeling more complicated. In the standard reflection spectroscopy regime, we find either an additional \texttt{xillverDCp} or a second \texttt{reltransDCp} could not fully describe the spectra, suggesting a different origin rather than the distant reflection. We preliminarily tested out two physically-driven models which turned out to both be promising: emission from the plunging region modeled by \texttt{bbody} \citep{fabian2020soft}, and returning radiation approximated by \texttt{relxillNS} \citep{connors2020evidence}. Other possibilities include extra Comptonization of the reflection spectrum in the corona \citep{wilkins2015driving,steiner2017self}, or a change in the disk thickness as the mass accretion rate rises. 

The best-fit parameter values are presented in Table~\ref{tab:spec_pars}. The iron abundance is close to the solar value at $1.24^{+0.10}_{-0.15}$. The disk temperature increases from $\sim0.3$~keV to $\sim0.6$~keV, which is expected in a hard-to-soft transition because the soft state spectrum becomes dominated by the thermal disk emission. The inner edge of the disk remains $<2$~$R_{\rm ISCO}$ in Epochs~1--4, but increases to $7.5^{+0.5}_{-0.9}$~$R_{\rm ISCO}$ in Epoch~5. The boost parameter enables the model to account for deviations from the model assumptions of an isotropically radiating source and point-like geometry, and it is $\sim2$ in Epochs~1--4, and becomes $3.0\pm1.5$ in Epoch~5 which causes the continuum and reflection components to overlap in Fig.~\ref{fig:eem_ratio}(e). However, we note that even though the best-fit value is larger than previous epochs, it is consistent within 90\% confidence. We notice several behaviors worth discussing: the inclination is pegged at the higher limit given by the binary inclination $i<81^\circ$ \citep{torres2020binary}, the ionization increases from $\sim2.5$ to $\sim4.5$. In fact, in another LMXB XTE~J1550--564, the inclination was also measured to be high when the source progressed to softer states \citep{connors2019inclination}. The reason behind these behaviors is very likely to be the same: the fit tries to accommodate the blue-shifted iron line which is prominent in Epochs~4--5, because a blue-shifted line could result from either a higher ionization or a higher inclination angle.

\begin{figure*}[htb!]
\centering
\includegraphics[width=0.8\linewidth]{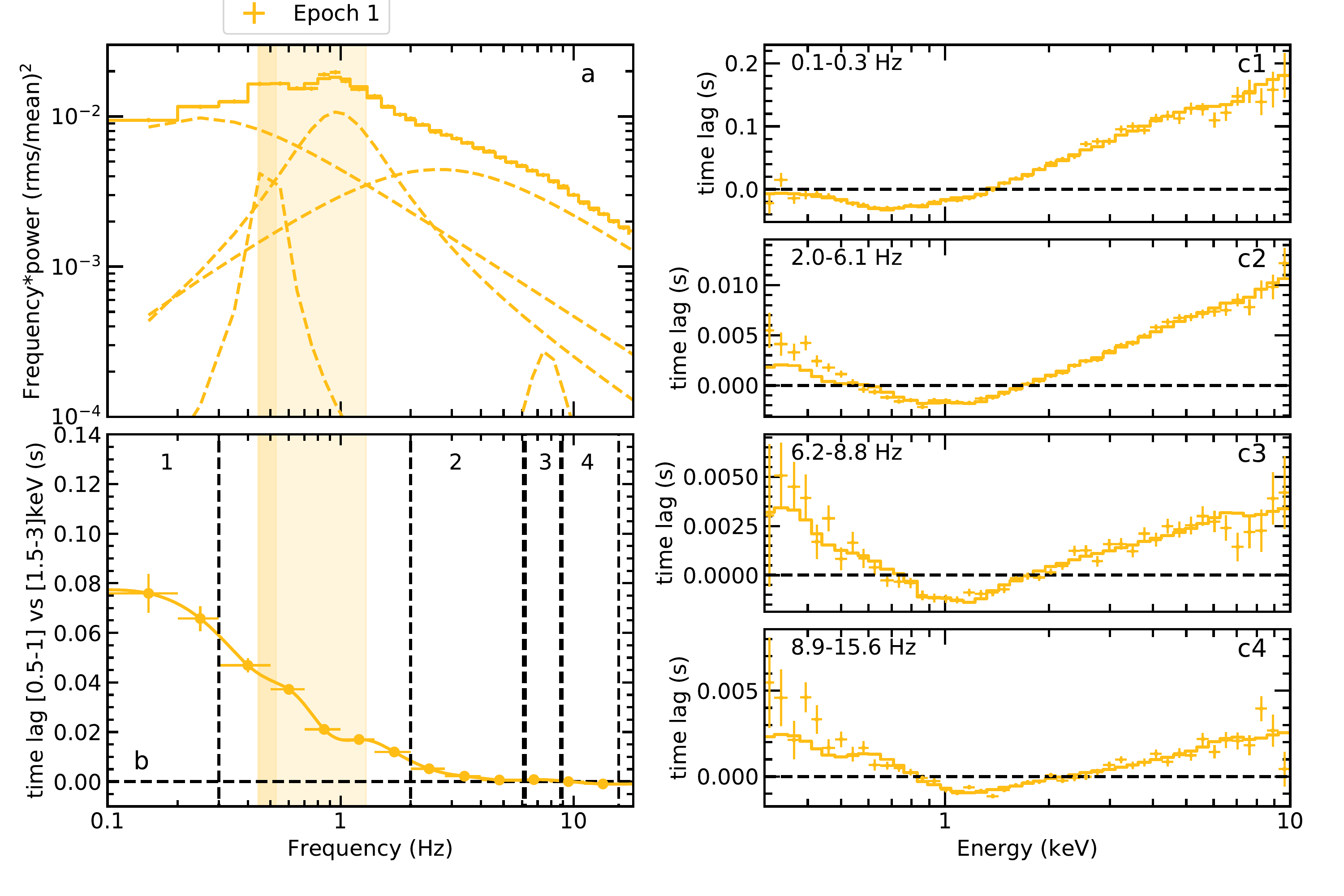}
\caption{The time lag fit result in Epoch~1. (a) The power spectral density, (b) the lag-frequency spectrum where the dashed lines represent the frequency ranges to extract and fit the lag-energy spectra (c1--c4), and the shaded regions are the QPO frequencies (central frequency $\pm{\rm HWHM}$) that we do not include in the time lag fits. The points in (b) are connected with a Bezier join to guide the eye. 
}
\label{fig:pds_lag_freq_lag_energy}
\end{figure*}

\subsection{Fitting the lag-energy spectra with the reverberation model} \label{lag_fit}

For each epoch, we fit simultaneously the lag-energy spectra in multiple frequency bands chosen to be outside of the central frequency $\pm$ half-width at half-maximum (HWHM) of QPOs and their (sub)harmonics fitted with the sum of several Lorentzians (\citealp{olive1998rxte}, see Fig.~\ref{fig:pds_lag_freq_lag_energy} for an example). The QPO frequencies and the frequency bands in which we extract the lag-energy spectra in all epochs are shown in Fig.~\ref{fig:qpo_freq_freq_bands}. The reason to avoid QPO frequencies in the fit is that we observe harder lags at the QPO frequencies in the lag-frequency spectra compared to adjacent frequency bins, which could be due to a more complicated mechanism producing the lags \citep{Ingram2016} rather than a combination of the pivoting power-law and the reverberation lag assumed in \reltrans\ (see Appendix for more details).

We employ the exact same model, \reltransDCp, used in the time-averaged spectral fit (Section~\ref{spec_fit})\footnote{To fit the lag-energy spectra accounting properly for the instrumental response, the \reltransDCp\ parameter ReIm is set to 6, and $f_{\rm min}$ and $f_{\rm max}$ are fixed to the lower and upper bounds of the frequency bands used to extract the lag spectra. The parameters of the pivoting power-law are free to vary in each frequency range.}. As it is computationally expensive to fit all the epochs simultaneously, we need to fit epochs individually. For consistency, we fix some global parameters to reasonable values for all epochs. Using the proper motion and the distance to the system from radio parallax measurement, $2.96\pm0.33$~kpc, the inclination of the jet to the line of sight is estimated to be $i=63\pm3^\circ$ \citep{atri2020radio}. Optical spectroscopy reveals a mass function of $f(M) = (M_1 \sin i)^3/(M_1+M_2)^2 = 5.18 \pm 0.15 M_\odot$, where $M_1$ and $M_2$ are the masses of the BH and its companion \citep{torres2019dynamical}; assuming the conservative range of mass ratio $0.03<q\equiv M_2/M_1<0.4$, the BH mass is $10\pm2$~$M_\odot$. Therefore, we fix the inclination to $i=63^\circ$, and the mass to $M=12$~$M_\odot$. The spin parameter is fixed at the maximal value of 0.998 to allow the inner disk radius to fit to any physically allowed value. The iron abundance is fixed at the solar value ($A_{\rm Fe}=1$). In the source direction, the Galactic column density is $N_H=1.3$--1.5 in units of $10^{21}$~cm$^{-2}$ \citep{bekhti2016hi4pi,Kalberla2005}, so we fix it at $1.3\times10^{21}$~cm$^{-2}$. In addition, we fix the coronal electron temperature $kT_{\rm e}=$100~keV because \nicer's energy coverage is insufficient to constrain it, and the boost parameter at 1, corresponding to a static, isotropically emitting lamppost corona. Guided by the prior expectation that the inner edge of the accretion disc extends to the ISCO by the time the source enters the soft state \citep{remillard2006x}, and to simplify our model, we set the inner edge of the disk, $R_{\rm in}$ to have an upper bound of $10$~$R_{\rm ISCO}$. As it is difficult to constrain the photon index $\Gamma$ with time lag fits alone, we only allow $\Gamma$ to vary between $\pm0.1$ around the corresponding best-fit value in the time-averaged energy spectral fit (Section~\ref{spec_fit}). 

\begin{table*}[htb!]
\caption{Best fit parameters when fitting the lag-energy spectra in several frequency ranges (outside the QPO frequencies) with the model \texttt{reltransDCp}.  \label{tab:tlag_orange_series}}
\centering
\begin{tabular}{ccccccc}\hline \hline
Parameter& Epoch 0  & Epoch 1 & Epoch 2 & Epoch 3 &Epoch 4 &Epoch 5\\
\hline
$i$ (degrees)&\multicolumn{6}{c}{$63$}\\
$A_{\rm Fe}$&\multicolumn{6}{c}{$1$}\\
$kT_{\rm e}$ (keV)&\multicolumn{6}{c}{$100$}\\
$a_*$&\multicolumn{6}{c}{$0.998$}\\
$M$ ($M_\odot$) & \multicolumn{6}{c}{$12$}\\
$N_H$ ($10^{21}$cm$^{-2}$)&  \multicolumn{6}{c}{$1.3$}\\
boost&\multicolumn{6}{c}{$1$}\\
$h$ ($R_{\rm g}$)& $42^{+1}_{-2}$ & $27^{+2}_{-3}$ & $33\pm5$ & $162^{+62}_{-45}$ & $114^{+34}_{-27}$ & $>335$ \\
$R_{\rm in}$ ($R_{\rm ISCO}$) & $4.0\pm 0.4$ & $6\pm 1$ & $6^{+2}_{-1}$ & $>5$ & $6^{(p)}_{-3}$ & $5^{+2}_{-1}$ \\
$\Gamma$& $1.57^{+0.02}_{-0.02}$ & $1.802^{+0.007}_{-0.002}$ & $1.800^{+0.004}_{(p)}$  & $2.01^{+0.07}_{-0.03}$ & $2.26^{(p)}_{-0.04}$  & $2.41^{+0.05}_{-0.01}$ \\
$\log \xi$ (erg$\cdot$cm$\cdot$s$^{-1}$)& $3.02^{+0.10}_{-0.06}$ & $2.09\pm0.06$ & $2.22\pm0.08$ & $1.40^{+0.39}_{-0.22}$ & $<3.1$ & $2.26^{+0.20}_{-0.52}$\\
$\log N_{\rm e}$ (cm$^{-3}$)& $19.74^{+0.03}_{-0.08}$ & $>19.99$ & $>19.93$ & $19.46^{+0.20}_{-0.36}$ & $>18.9$ & $18.59^{+0.28}_{-0.15}$\\
$\chi^2$/d.o.f.& $857.0/205$ & $270.2/163$ & $350.9/205$ & $223.4/205$ & $228.8/163$ & $129.1/79$\\
\hline
\end{tabular}
\\
\raggedright{\textbf{Notes.} \\
Errors are at 90\% confidence level and statistical only. Parameters without uncertainties were fixed at the quoted value. The index $(p)$ means that the uncertainty is pegged at the bound allowed for the parameter. Guided by the prior that the inner edge of the accretion disc extends to the ISCO by the time the source enters the soft state \citep{remillard2006x}, and to simplify our model, we set the inner edge of the disk, $R_{\rm in}$ to have an upper bound of 10~$R_{\rm ISCO}$. It is pegged at that limit in Epochs~3--4. The photon index, $\Gamma$, is set to vary between $\pm0.1$ around the best-fit value in the time-averaged energy spectral fit in Section~\ref{spec_fit} (1.8, 1.9, 2.0, 2.2, 2.4 for Epochs~1--5; 1.6 for Epoch~0, which corresponds to Epoch~2 in \citealp{buisson2019maxi}), and is pegged at the lower and upper bounds in Epochs~2 and 4 respectively.} 
\end{table*}

The resulting parameter values in all epochs are shown in Table~\ref{tab:tlag_orange_series}. The power spectral density, the frequency ranges used to extract the lag-energy spectra and then fitted, the best-fitted \reltransDCp\ model and lag-energy spectra for Epoch~1, are shown in Fig.~\ref{fig:pds_lag_freq_lag_energy}, as an example. We find \reltransDCp\ can describe the two types of time lags very well: the low-frequency hard lag, and the high-frequency soft/reverberation lag. At low frequencies, the reverberation lag contributes a dip at the iron line energy \citep{mastroserio2018multi}. The inner edge of the disk, $R_{\rm in}$ is consistent at $\sim6$~$R_{\rm ISCO}\sim7$~$R_{\rm g}$ (when $a_*=0.998$) where $R_{\rm ISCO}$ is the radius of the inner most stable circular orbit (ISCO), whose radius is a function of the spin, decreasing from 6~$R_{\rm g}$ to 1.23~$R_{\rm g}$ when the spin increases from $a_*=0$ to 0.998. With reflection spectroscopy in the hard state ($R_{\rm in}\sim5.3$~$R_{\rm g}$; \citealp{buisson2019maxi}) and continuum fitting method in the soft state \citep{zhao2020estimating, guan2020physical}, \maxij1820\ is expected to be a slow-spinning black hole with $a_*\sim0-0.2$. Therefore, our measured disk truncation level is consistent with a disk extending very close to the ISCO radius during the state transition. We also note that the disk electron density tends to peg at the upper limit of the model, $10^{20}$~cm$^{-3}$, confirming the necessity for BHB modeling to go beyond the usually tabulated value. 

\begin{figure}[htb!]
\centering
\includegraphics[width=1.\linewidth]{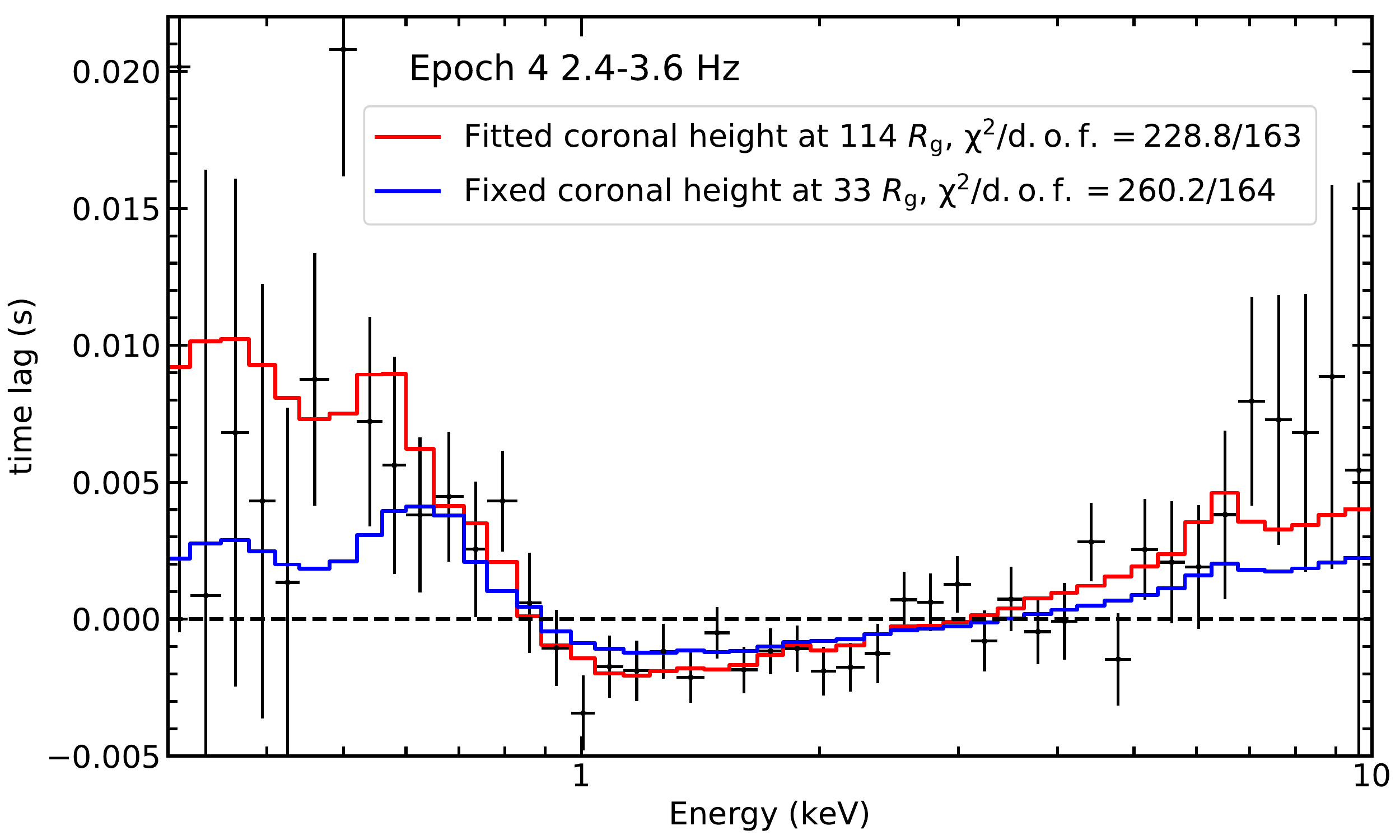}
\caption{The lag-energy spectrum and the best-fit models in Epoch~4, and frequency band of 2.4--3.6~Hz. The fit is much better when the coronal height is fitted at $\sim100$~$R_{\rm g}$ (red) than when it is fixed at the value in Epoch~2 ($33$~$R_{\rm g}$, blue). The $\chi^2$ values quoted are for the simultaneous fit of the 4 frequency bands.
}
\label{fig:compare_h}
\end{figure}

The key result is that the fitted coronal height decreases from $\sim40$~$R_{\rm g}$ at the beginning of bright hard state (Epoch~0) to $\sim30$~$R_{\rm g}$ at the end of hard state (Epochs~1--2), and then increases to $\sim100-200$~$R_{\rm g}$ when the transition begins (Epochs~3--4), and eventually to $\gtrsim300$~$R_{\rm g}$ in Epoch~5 just before the jet line. This trend suggests the corona is contracting in the hard state, and then is expanding in the hard-to-soft transition, which is consistent with the qualitative argument by tracking the evolution of reverberation lag frequency in Section~\ref{phenomena}. The increase in coronal height is strongly suggested by the lag spectra in our reverberation model because the statistics are much worse if the coronal height is fixed at the value found in Epoch~2 at the end of hard state ($33$~$R_{\rm g}$). The $\chi^2$ increases by 11, 31, and 25 with 1 more d.o.f. in Epochs~3--5 respectively. The largest difference in fitted models lies in both low energies below $\sim0.8$~keV, and also energies above $\sim3$~keV (e.g., see Fig.~\ref{fig:compare_h}). We also want to emphasize that even though some parameters are fixed, the trend in the coronal height persists in all the fitting trials with different setups, which is reasonable because within the reverberation model, the reverberation lag frequency fundamentally correlates with the size of the corona-disk emitting region. We note that the fit statistics are worse in Epochs~0--2 compared to Epochs~3--5 likely because the signal-to-noise ratios of lags are higher in the early epochs, while the variability decreases rapidly when the source approaches the jet line/soft intermediate state. There are common residuals at low energies $<1$~keV in all epochs, which is similar to what is shown in Fig.~\ref{fig:pds_lag_freq_lag_energy}, and could suggest some other contributions to the lags (e.g. an additional thermalization time lag; see the discussion in Section~\ref{discussion}).



\begin{figure*}[htb!]
\centering
\includegraphics[width=0.9\linewidth]{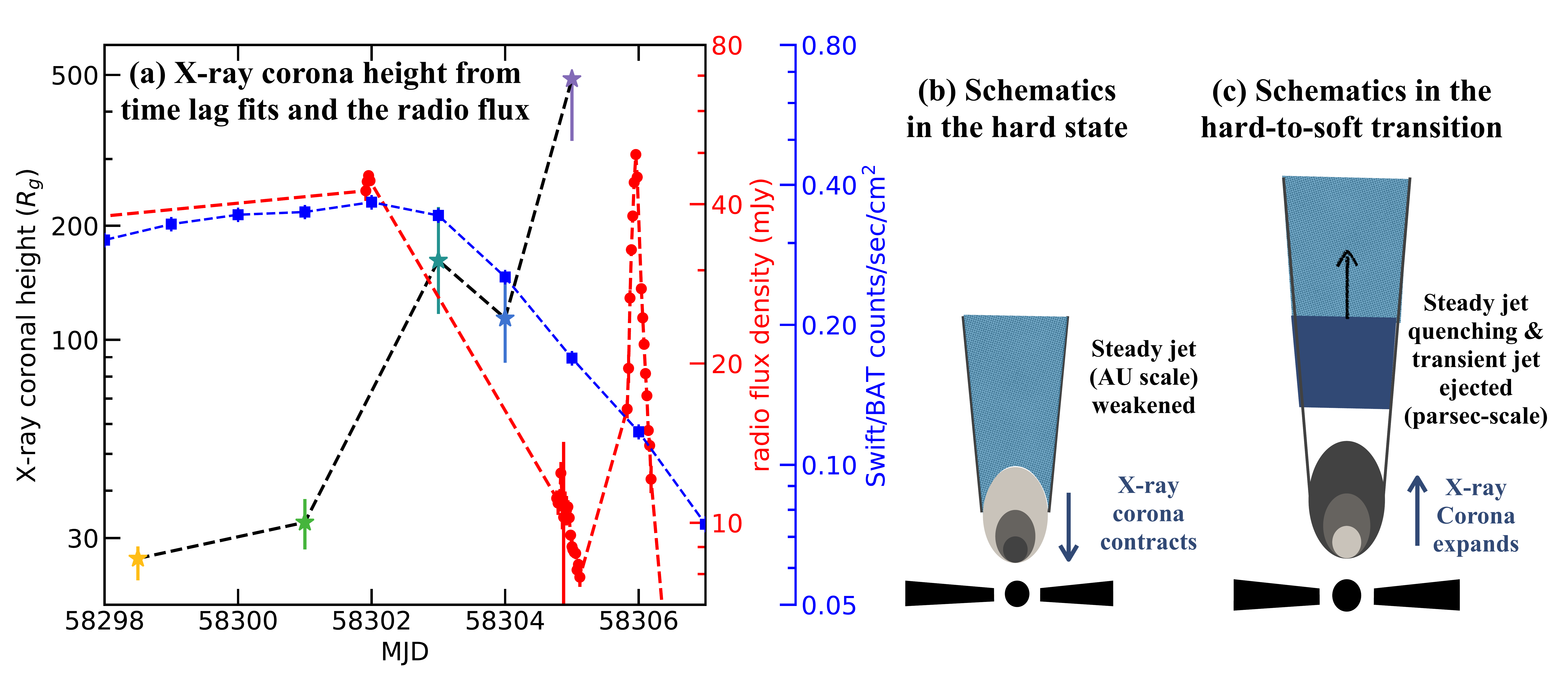}
\caption{Close relationship between the X-ray corona and radio jet behavior. (\textit{a}) The X-ray coronal height from time lag fits (rainbow colors corresponding to those in Fig.~\ref{fig:hid}, joined with the black dash line), the radio flux density from AMI (red) and the \swift/BAT count rate in 15-50~keV (blue), as functions of MJD time. The coronal height refers to the left black axis, the radio flux density and the \swift/BAT light curve refer to the right axis in red and blue. (\textit{b}) In the hard state, the X-ray corona contracts while the steady compact jet is weakened in the radio. (\textit{c}) In the hard-to-soft state transition, the X-ray corona expands, and there launched a ballistic transient jet (superposed on the quenching steady jet) several days lagging behind the X-ray change.}
\label{fig:radio_h}
\end{figure*}

\subsection{Comparison between time-lags and flux spectra, and lessons learned}

Interestingly, some physical parameter estimates from spectral and time-lag fits do not agree. The most dramatic discrepancy lies in the coronal height during the hard-to-soft state transition, where the fit to the flux spectra finds a fairly low coronal height, of the order of 10~$R_{\rm g}$ with a decreasing trend, while time-lag fit results in an increase from tens of $R_{\rm g}$ to hundreds. One plausible explanation is that the corona is more vertically extended in the transition, forming the base of the transient jet. This explanation is also consistent with the fact that the coronal heights from the time-lag and flux spectral fit in Epochs~1 and 2 agree very well, because the corona is more compact at the beginning of the transition.
Once the corona is more extended, fits with a single lamppost source are expected to yield a larger source height from the lag spectrum than from the flux spectrum, because the lower regions of the corona dominate the reflection emissivity but reflection from the upper regions contributes a larger light-crossing delay.

It has been commonly seen from high signal-to-noise energy spectra that besides the relativistically broadened iron line, a narrow core of the line is required that can be described by an unblurred reflection component.  \citet{garcia20192017} recently found that an approximation of the corona by two lamppost illuminators offer a better description of the spectra from GX~339$-$4, i.e., the distant reflector still has some relativistic effects encoded. The dual-lamppost reflection model is also successfully adopted in \citet{buisson2019maxi} when analyzing \maxij1820\ spectra. The upper lamppost height is found to decrease in the hard state, perfectly consistent with the qualitative result with reverberation lag frequencies \citep{kara2019corona}. However, it is not that straightforward to add a second lamppost source in the time lag fits, because that would require a proper combination of two response functions in the complex plane and a re-calculation of the ionization profile due to the dual-lamppost geometry illumination.

\section{Discussion} \label{discussion}

With the extraordinary \nicer\ dataset of \maxij1820\ in the bright hard-to-soft transition, we are able to find that the reverberation lag frequency decreases in the transition, with an increasing amplitude, which indicates a larger emitting region that causes reverberation, likely due to an expanding corona. With the reverberation model \reltransDCp, we are able to quantitatively measure the coronal height, by fitting both the lag-energy spectrum and the time-averaged energy spectrum. In the time lag fits, the X-ray coronal height increases from several tens to several hundred $R_{\rm g}$ several days before the radio flare, consistent with the qualitative trend seen in reverberation lags. An expanding or ejected corona during a state transition may explain earlier RXTE observations of the high-mass X-ray binary Cyg~X-1, where the low-frequency hard lags were seen to increase during the transition \citep{pottschmidt2000temporal,grinberg2013long}. We want to stress that to our knowledge, this is the first BHB in which the increase of reverberation/soft lags is clearly observed.


Next, we will discuss this result in the context of the radio jet. Previous work suggested that the corona contracts during the hard state \citep{kara2019corona}, when the steady jet becomes weaker. This may suggest that the jet luminosity is related to the vertical extent of the X-ray corona, and would favor the hypothesis that the corona is the base of a jet. Now, we have extended this methodology to the hard-to-soft state transition. Our measured X-ray coronal heights from fitting the time lags, and the radio flux density from the Arcminute Microkelvin Imager (AMI; \citealp{zwart2008arcminute,hickish2018digital}) during the transition are presented in Fig.~\ref{fig:radio_h}~(a). The radio flux density indicates a weakened steady jet, following the trend in the hard state, followed by a powerful transient jet in the transition that features ballistic ejections \citep{bright2020extremely}, before the source transitions to the soft state.

We propose that the increase in coronal height inferred from our reverberation analysis coincides with the launch of the blob that $\sim5$~days later causes a radio flare when it reaches a large enough scale to become radio bright (Fig.~\ref{fig:radio_h}~b, c). We note that due to the low-cadence radio monitoring between MJD~58302 and 58304, we cannot rule out that there was another earlier flare that was missed, and thus the time delay between corona expansion and radio flare would be less than 5~days. If the blob is launched when the corona is most compact (MJD $\sim 58298$) and travels at $\sim 0.9 c$ \citep{bright2020extremely}, then a delay of $\sim 5$ days until a radio flare
would suggest that the radio-emitting blob is at $10^9$~$R_{\rm g}$. Unfortunately, there have been no published radio measurements of the synchrotron break to independently estimate the radius and distance of the jet ejecta, but this distance is reasonable, when compared to the BHB MAXI~J1535--571 where the emission height above the BH is $\sim 10^8$~$R_{\rm g}$ \citep{russell2020rapid}. We also note that recent Chandra imaging showed a transient X-ray jet ejecta at the time of the radio flare, inferring that the ejecta was a cone of $1.5\times 10^4$~AU by $7.7\times 10^3$~AU \citep{espinasse2020relativistic}. As 1~AU$\sim10^7$~$R_{\rm g}$ for a 10~$M_\odot$ BH, this would lead to a vertical extent of the transient jet emitting in X-ray of $\sim10^{11}$~$R_{\rm g}$, and the light travel distance of several days is always below this upper limit. Therefore, it is plausible that the expansion of X-ray corona is accompanied with a jet knot which propagates downstream (moves higher up) with the jet material when approaching the time when the radio flare was observed. This interpretation is also suggested to explain the increased hard lag amplitude during the transition in the high mass X-ray binary Cyg~X-1 \citep{pottschmidt2000temporal}. It is consistent the physical picture presented in \citet{russell2019disk}, where the soft X-ray flux is observed to increase just before the radio flare in another LMXB MAXI~J1535--571, because the X-ray flare could result from inverse Compton scattering of the synchrotron radiation from the knot after it was ejected \citep{russell2019disk}. Finally, this behavior may be analogous to observations in some high luminosity AGN, where X-ray flares and iron line narrowing precede radio ejection events (\citealp{marscher2008inner,lohfink2013,king2016discrete}). In this scenario, the source of the hard power-law emission is more likely to arise from synchrotron emission in addition to Thermal Comptonization because the optical depth along the jet axis ($\tau\propto N_j/r$ where $N_j$ is the power injected in the jet, and $r$ is the radius of the jet) drops quickly \citep{lucchini2021correlating}, while efficient thermal Comptonization requires a mildly optically thin medium ($\tau\sim0.1-1$).

We also notice that the corona is found to be physically and radiatively compact within several tens of $R_{\rm g}$ from microlensing in lensed quasars \citep{chartas2016gravitational}, X-ray reverberation mapping in Seyfert galaxies \citep{kara2016global}, and X-ray eclipses of the corona (e.g. \citealt{gallo2020}). Therefore, the exact values of the coronal height (especially the ones at hundreds $R_{\rm g}$ in time lag fits) need to be taken with caveats, such as possible contribution from thermalization time in the disk \citep{garcia2013time} which we will implement in \reltrans\ the near future, or even the effects from returning radiation that are shown to increase the reverberation lags by $\sim50$\% \citep{wilkins2020returning}. With all this in mind, we stress that within the reverberation model, the relative trend of increasing coronal height during the state transition comes fundamentally from the change in the reverberation lag frequency.

\section{Summary} \label{summary}

We have analyzed five epochs of \maxij1820\ in the hard-to-soft transition using \nicer, to study the spectral-timing evolution of the system. Our major findings are as follows. 

\begin{enumerate}
\item{During the hard-to-soft transition, the frequency range where the reverberation lag dominates decreases, and reverberation lag amplitude increases, suggesting a growing X-ray emitting region, possibly due to an expanding corona.}
\item{From the flux spectral fit, the inner edge of disk does not decrease. This result challenges the state transition model of a collapsed hot inner flow.}
\item{By fitting the lag-energy spectra in a set of Fourier frequency ranges with the reverberation model \reltransDCp, we measure the coronal height to be anomalously large. The height of the corona increases $\sim5$~days before a radio jet ejection event.}
\item{We propose that the correlation between the X-ray corona height and the radio jet behavior can be explained with a model where the X-ray corona is the base of the jet which in its final moments before transitioning to the soft state, is ejected.}
\end{enumerate}

\bigskip
While this work was being finalized, we became aware of \cite{de2021inner}, showing the decrease in the observed soft lag frequency during the state transition, corroborating this aspect of our work. JW, GM, EK, JAG, AI and ACF acknowledge the International Space Science Institute (ISSI) for hosting the Workshop ``Sombreros and lampposts: The Geometry of Accretion onto Black Holes." JW, GM, EK and JAG acknowledge support from NASA~ADAP grant 80NSSC17K0515. JAG thanks support from the Alexander von Humboldt Foundation. AI and DB acknowledge support from the Royal Society.
\appendix

\setcounter{figure}{0}
\renewcommand{\thefigure}{A\arabic{figure}}

\begin{figure}[htb!]
\centering
\includegraphics[width=0.6\linewidth]{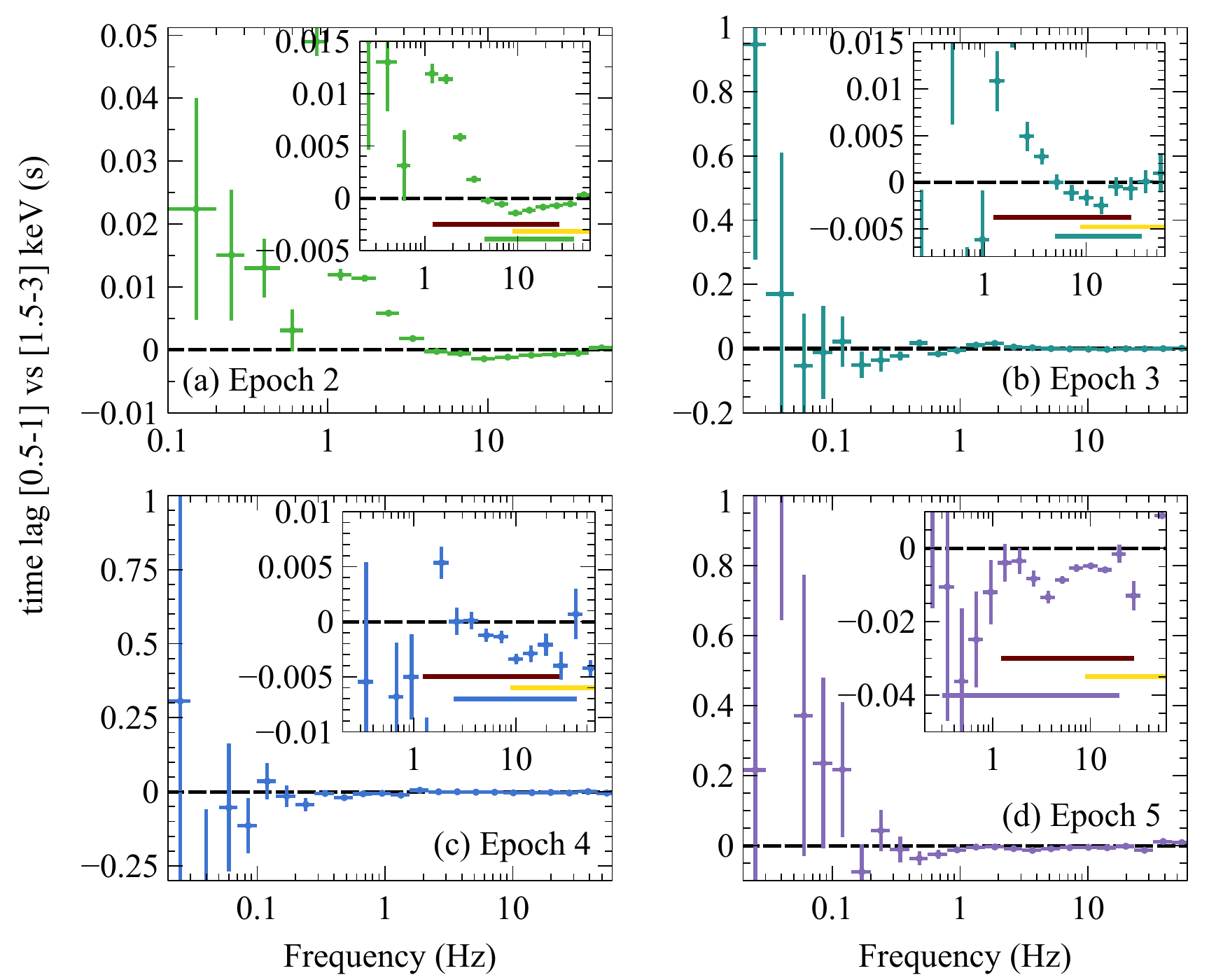}
\caption{The decrease of reverberation lag frequency can also be seen in individual epochs. The horizontal lines in the inserted plot represent the frequency ranges of soft lags, and the red and yellow frequency ranges are for Epochs~0 and 1 at the start and end of hard state, which serve as benchmarks. \label{fig:individual_tlag}}
\end{figure}

\begin{figure}[htb!]
\centering
\includegraphics[width=0.6\linewidth]{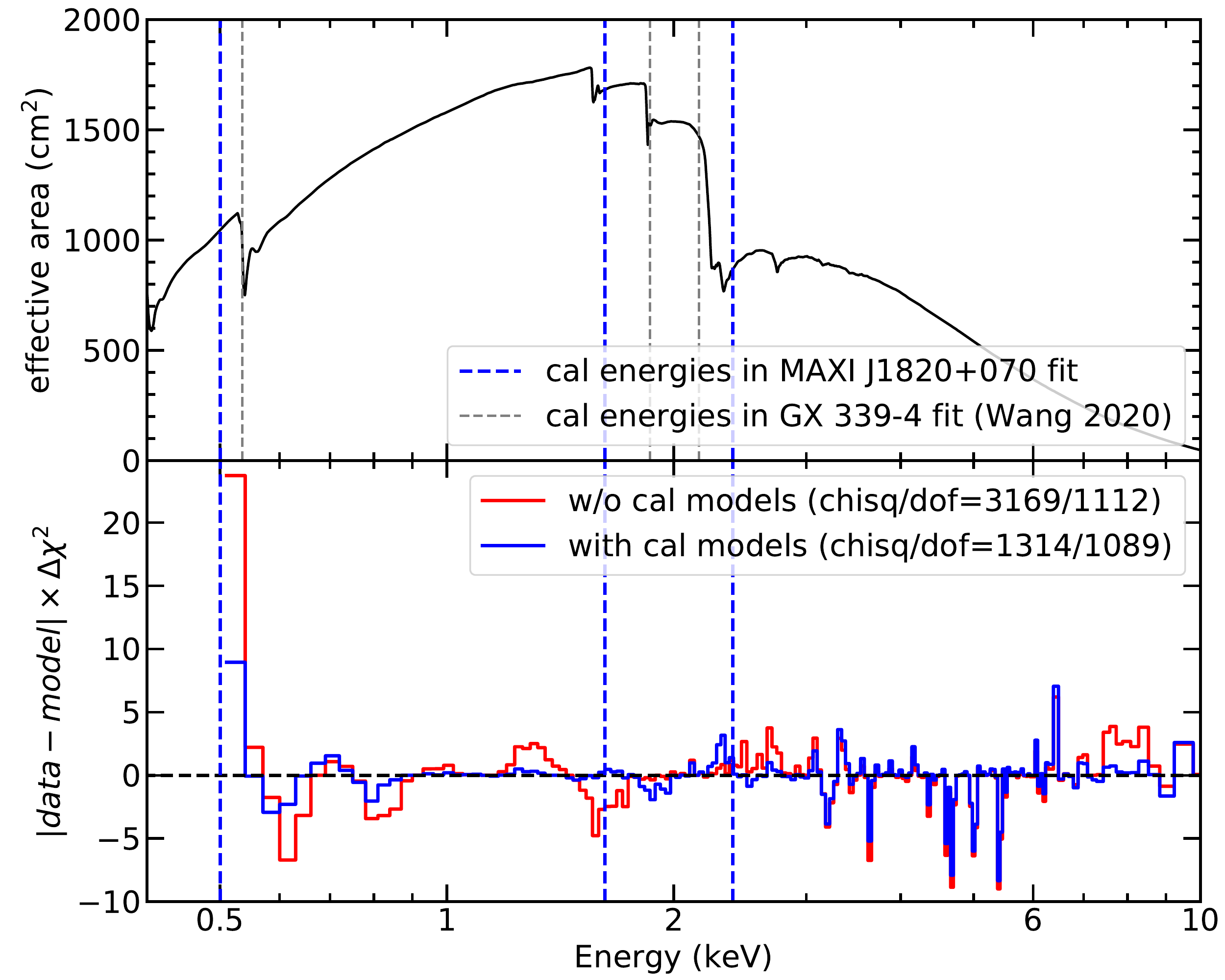}
\caption{(\textit{Upper}) The effective area of \nicer, with the fitted characteristic energies of phenomenological calibration models in \citet{wang2020relativistic} and this work. (\textit{Lower}) The residuals for \nicer\ data in Epoch~3 from simultaneous fits with or without the phenomenological calibration models. The $\chi^2$ values quoted are for the simultaneous fit of the 5 epochs. \label{fig:cal_model}
}
\label{fig:arf_chi}
\end{figure}

\begin{figure}[htb!]
\centering
\includegraphics[width=0.6\linewidth]{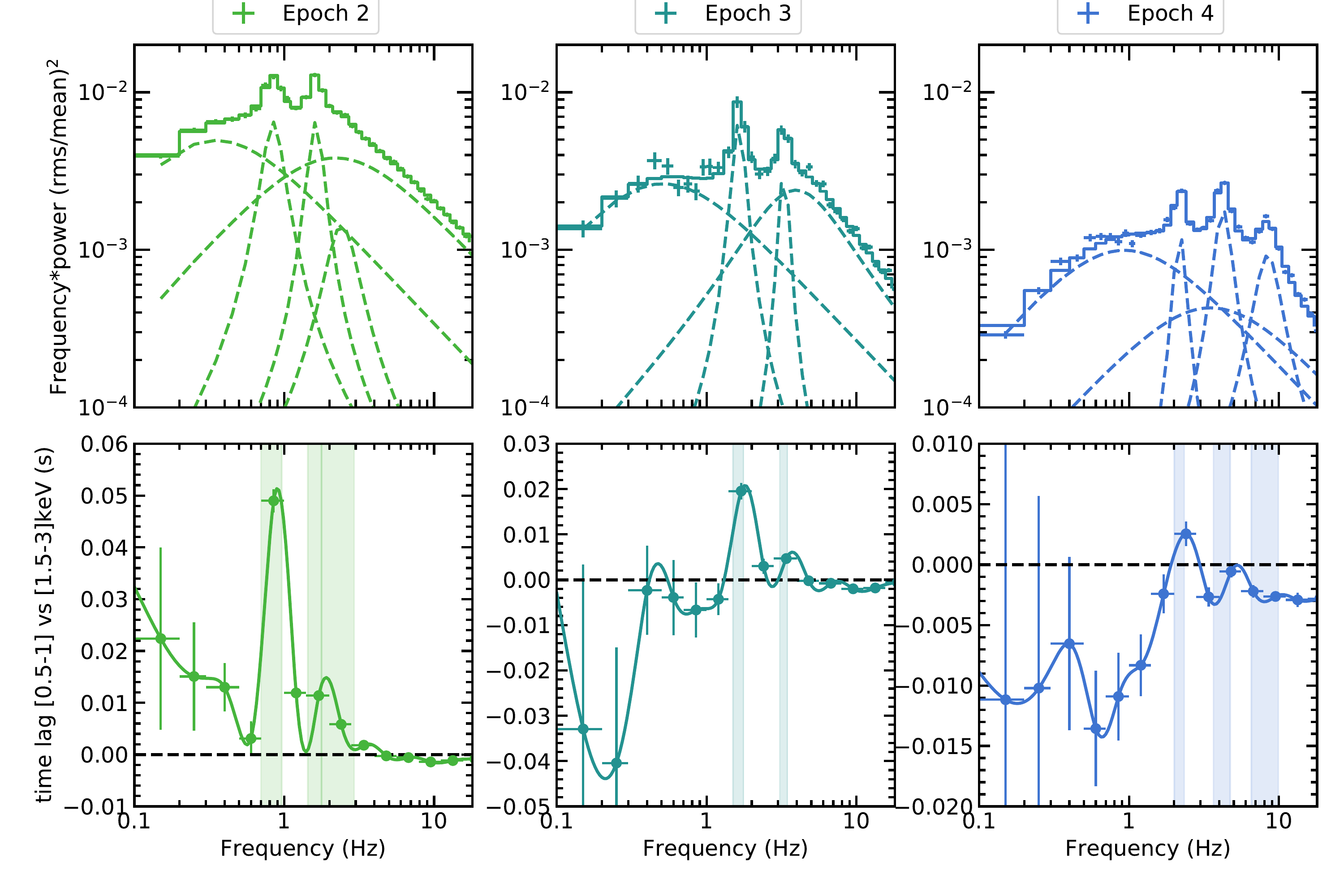}
\caption{(\textit{Upper}) The power spectral densities and the fitted Lorentzian components. (\textit{Lower}) The lag-frequency spectra, where the shaded regions represent the QPO frequencies (central frequency $\pm{\rm HWHM}$). We find ``harder" lags at the frequencies of type-C QPOs and their (sub)harmonics than at neighbouring frequencies. \label{fig:qpo_lag}
}
\end{figure}

For the epochs during the transition, We find ``harder" lags at the frequencies of type-C QPOs and their (sub)harmonics than at neighbouring frequencies (Fig.~\ref{fig:qpo_lag}). This suggests a very likely different or at least more complicated mechanism to produce the time lags involving geometric effects such as the Lense-Thirring procession \citep{stella1997lense}, rather than a combination of the pivoting power-law and reverberation as considered in \reltrans. The phase lags at the type-C QPO frequencies have been systematically investigated in \citet{van2016inclination, zhang2020systematic}, but a similar feature has not been reported. Perhaps this is because those observations were mostly of type-C QPOs in the hard state while we find that such hard lags are specific to type-C QPOs observed during the relatively short-lived transition and are not seen as clearly in the hard state.

\begin{figure}[htb!]
\centering
\includegraphics[width=0.6\linewidth]{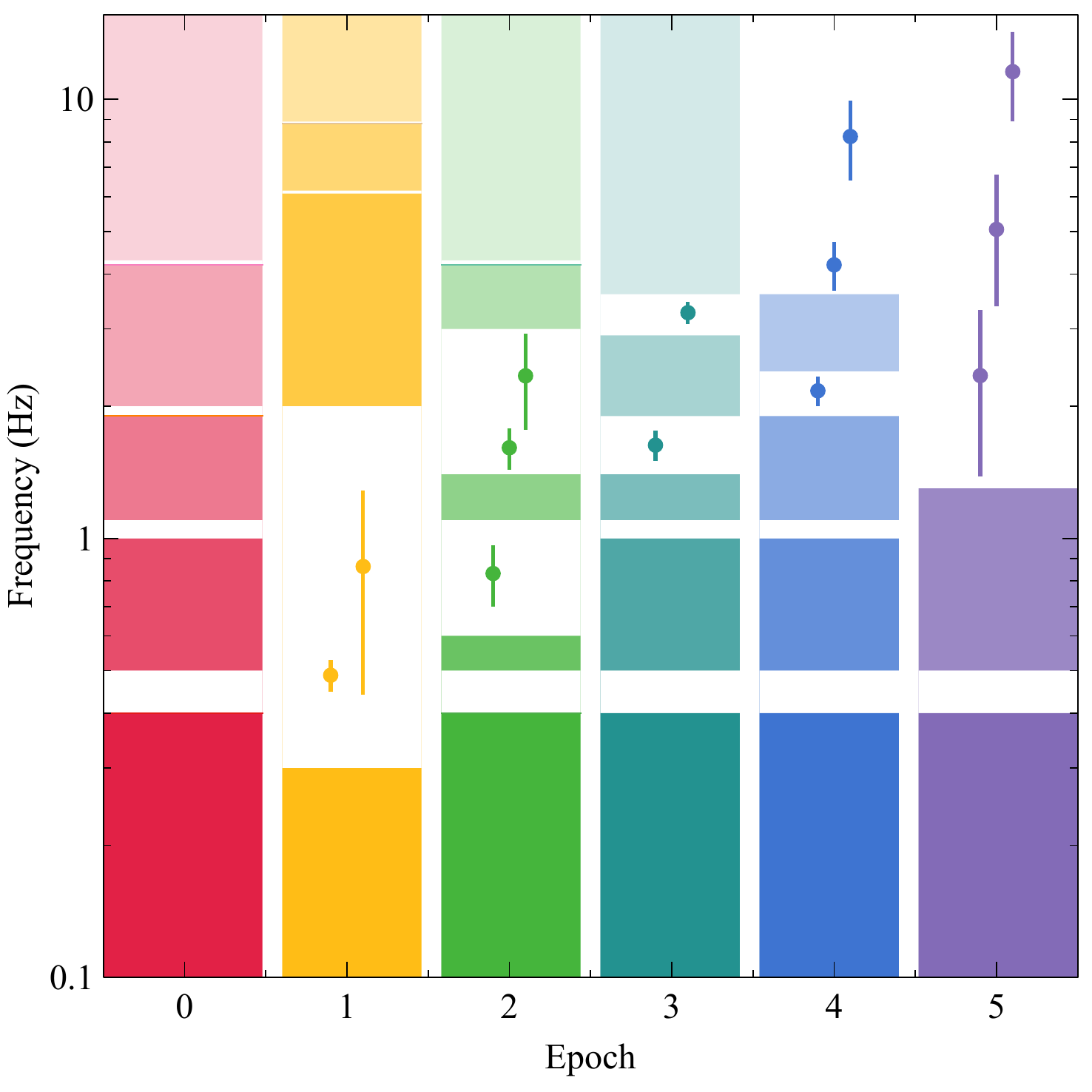}
\caption{The QPO frequencies (central frequency $\pm{\rm HWHM}$), and the frequency bands outside of those QPO frequencies in which we extract the lag-energy spectra and fit with \reltransDCp\ in each epoch. The data points representing QPO frequencies are shifted in x-axis for clarity.
}
\label{fig:qpo_freq_freq_bands}
\end{figure}

\bibliographystyle{apj}
\bibliography{draft3}
\end{document}